\documentclass[prl,noshowpacs,superscriptaddress,twocolumn]{revtex4-1}
\usepackage{amsmath}
\usepackage{amsfonts}
\usepackage{graphicx}
\usepackage{multirow}
\usepackage{chngcntr}
\usepackage{bm}

\usepackage{color}
\usepackage{ulem}
\usepackage[rgb,dvipsnames]{xcolor}

\usepackage{natbib}
\usepackage[final=true]{hyperref}
\hypersetup{
    colorlinks=true,
    linkcolor=cyan,
    filecolor=magenta,      
    urlcolor=cyan,
    citecolor=blue,    
}

\begin{document}

\title{Simultaneous creation of multiple vortex-antivortex pairs in momentum space in photonic lattices}

%\title{The momentum space vortices in photonic graphene and hexagonal lattices}

\author{Feng Li}
\affiliation{Key Laboratory for Physical Electronics and Devices of the Ministry of Education \& Shaanxi Key Lab of Information Photonic Technique, School of Electronic Science and Engineering, Faculty of Electronics and Information, Xi'an Jiaotong University, Xi'an 710049, China}
%\email{felix831204@xjtu.edu.cn}

\author{S.~V.~Koniakhin}
\email{kon@mail.ioffe.ru}
\affiliation{Center for Theoretical Physics of Complex Systems, Institute for Basic Science (IBS), Daejeon 34126, Republic of Korea}

\author{A.V. Nalitov}
\affiliation{Faculty of Science and Engineering, University of Wolverhampton, Wulfruna Street, Wolverhampton WV1 1LY, United Kingdom}
\affiliation{ITMO University, St. Petersburg 197101, Russia}

\author{E. Cherotchenko}
\affiliation{Ioffe Institute, 194021 St. Petersburg, Russia}

\author{D.D. Solnyshkov}
\affiliation{Institut Pascal, PHOTON-N2, Universit\'e Clermont Auvergne, CNRS, Clermont INP,  F-63000 Clermont-Ferrand, France}
\affiliation{Institut Universitaire de France (IUF), 75231 Paris, France}

\author{G. Malpuech}
\affiliation{Institut Pascal, PHOTON-N2, Universit\'e Clermont Auvergne, CNRS, Clermont INP,  F-63000 Clermont-Ferrand, France}

\author{Min Xiao}
\affiliation{Department of Physics, University of Arkansas, Fayetteville, Arkansas, 72701, USA}
\affiliation{National Laboratory of Solid State Microstructures and School of Physics, Nanjing University, Nanjing 210093, China}

\author{Yanpeng Zhang}
\affiliation{Key Laboratory for Physical Electronics and Devices of the Ministry of Education \& Shaanxi Key Lab of Information Photonic Technique, School of Electronic Science and Engineering, Faculty of Electronics and Information, Xi'an Jiaotong University, Xi'an 710049, China}

\author{Zhaoyang Zhang}
\email{zhyzhang@xjtu.edu.cn}
\affiliation{Key Laboratory for Physical Electronics and Devices of the Ministry of Education \& Shaanxi Key Lab of Information Photonic Technique, School of Electronic Science and Engineering, Faculty of Electronics and Information, Xi'an Jiaotong University, Xi'an 710049, China}

\begin{abstract}

Engineering of the orbital angular momentum (OAM) of light due to interaction with photonic lattices reveals rich physics and motivates potential applications. We report the experimental creation of regularly-distributed quantized vortex arrays in momentum space by probing the honeycomb and hexagonal photonic lattices with a single focused Gaussian beam. For the honeycomb lattice, the vortices are associated with Dirac points and mimic the Berry curvature sources. However, we show that the resulting spatial patterns of vortices are strongly defined by the symmetry of the wave packet evolving in the optical lattice but not by lattice topological properties. Our findings reveal the underlying physics by connecting the symmetry and OAM conversion, and provide a simple and efficient method to create regularly-distributed multiple vortices by unstructured light.
\end{abstract}

\maketitle

A vortex is a fundamental topological structure associated with the circulation loops in various vector fields. Quantized vortices present in quantum fluids were discovered as elementary topologically nontrivial excitations of superfluid helium \cite{onsager1949statistical} and superconductors \cite{abrikosov1957magnetic}. In scalar quantum fields, the existence of vortices is allowed by the gauge invariance of the wave functions (WFs) related to the symmetry to phase shifts of $2\pi n$ ($n\in\mathcal{Z}$). The phase ambiguity at the vortex core requires a vanishing WF density, which is a vivid observable signature.
Another manifestation is diverging momentum curl, which results in finite energy and angular momentum contribution associated with a lower-dimensional vortex core. Importantly, in many-body systems of interacting particles, described with the Gross-Pitaevskii equation, quantum vortices are stable solutions whose spatio-temporal dynamics governs the quantum turbulence \cite{bradley2012energy,Gauthier2019,Johnstone2019,koniakhin20202d}. Stable and well-defined quantum vortices are also present in driven-dissipative systems, such as exciton-polariton interacting quantum fluids \cite{Carusotto2013,lagoudakis2008quantized,koniakhin2019stationary,claude2020taming}, where they are responsible for the Berezinskii–Kosterlitz–Thouless transition \cite{Caputo2018}.
The polaritonic spin-orbit coupling due to TE-TM splitting  \cite{kavokin2005optical,koniakhin2021topological} allows new topologically nontrivial excitations such as half-integer vortices \cite{Rubo2007} and polarization vortices \cite{Dufferwiel2015} in semiconductor microcavities.

Along with the real-space vortices, similar topological configurations exist in the reciprocal (momentum) space. Such vortices may stem from the nontrivial topology in band structures, manifesting Berry curvature sources. In graphene, momentum-space vortices induced by pseudospin exist at Dirac points \cite{bena2009remarks,fuchs2010topological,lim2015geometry} and their topological charges have opposite signs in the two nonequivalent valleys in the valence and conduction bands. In general, pseudospin (sublattice) and valley (K and K') degrees of freedom lead to rich optical behavior of artificial graphene and related photonic lattices for orbital angular momentum (OAM) managing. In photonic honeycomb and Lieb lattices, pseudospin-orbit interaction allowed obtaining the conversion of pseudospin and OAM \cite{liu2020universal} in the angular momentum of the output signal. At the same time, valley-selective excitation by structured field in photonic graphene leads to the emergence of vortices of topological charge with arbitrary sign in the Bragg-reflected component in the complementary Dirac points \cite{song2019valley}. Polarization vortices in momentum space were reported in topological honeycomb and Lieb plasmonic lattices \cite{zhang2018observation}. The strong spin-orbit coupling was employed to give rise to the appearance of spin antivortices in momentum space in atomically-thin metals with graphene-like structure \cite{yang2022momentum}. 

Recently, the artificial photonic lattices based on coherent atomic vapors with refraction index modulated by electromagnetically-induced transparency (EIT) \cite{gea1995electromagnetically} have shown their advantages for analogue physics: the real-time tuning of the lattice potential that allows to study the dynamics of the wave packets. This technique boosted the investigation and visualization of angle-dependent Klein tunneling \cite{zhang2021angular} and edge-state solitons \cite{Zhang2020}, as well as the observation of particle-like dynamics of the vortex cores formed during the wave packet evolution in photonic graphene \cite{zhang2019particlelike}. So far, the conversion between OAM and the pseudospin via the Dirac point involves complicated beam structures to ensure the excitaion of a single pseudospin of the lattice, which generates one specific vortex corresponding to the excited pseudospin \cite{liu2020universal,song2019valley,zhang2019particlelike,Zhang2020Optica}. Therefore, it is essential to find alternative mechanisms for creating vortices with simpler schemes, better efficiency and richer physics. Meanwhile, as all the above-mentioned systems involve geometrically-symmetric lattices which display topological features, it is not clear whether topology is a necessity for the generation of vortices, or the vortices may simply originate from the lattice symmetry. 

In this article, we report the creation of momentum-space vortices in atomic-vapor-cell-based honeycomb and hexagonal photonic lattices excited by a non-structured Gaussian probe beam, and demonstrate that vortices in reciprocal space may emerge even in topologically trivial systems due to the interplay of symmetries of the excitation and the lattice. Such scheme allows easy generation of multiple vortex-antivortex pairs uniformly distributed in the momentum space upon a single excitation. Moreover, the resulting OAM textures can be tuned via the probe symmetry. Our study reveals the underlying physics between symmetry and vortex generation, and indicates that  topology does not necessarily play an important role in such processes.

\begin{figure}
\includegraphics[width=0.5\textwidth]{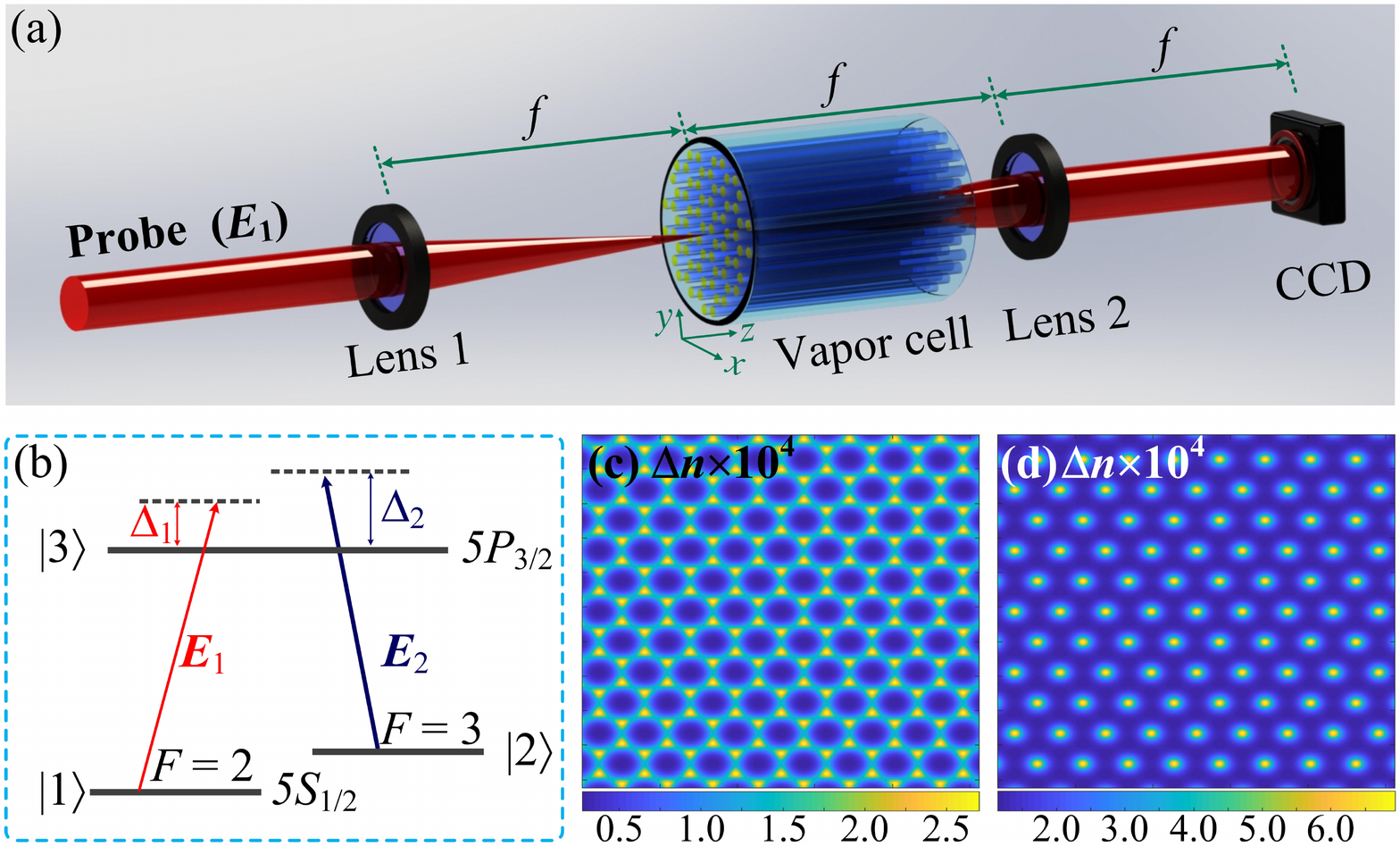}
\caption{\label{fig_exp_setup} Experimental Schemes. (a) Illustrative picture of the experimental setup. The focus lengths of the two lenses are both $f=150$~mm. (b) The three-level $^{85}$Rb atomic configuration excited by the probe field \bm{$E_1$} and the hexagonal coupling field \bm{$E_2$}. (c) and (d) Calculated spatial distribution of refractive index for $\Delta_1=-80$~MHz (c) and $\Delta_1=-190$~MHz (d) with $\Delta_2=-100$~MHz, resulting in honeycomb and hexagonal lattices, respectively.}
\end{figure}

The photonic lattice is created by engineering the susceptibility via EIT effect in a  $^{85}$Rb vapor cell \cite{gea1995electromagnetically, zhang2019particlelike}. As sketched in  Fig.~\ref{fig_exp_setup}(a), the coupling field \bm{$E_2$} is constructed by the interference of three light beams from the same laser, forming a 2-dimensional (2D) hexagonal pattern in the $x$-$y$ plane propagating along $z$. For simplicity, the optical path of \bm{$E_2$} is not shown in the graph, with technical details already introduced in Ref.~\cite{zhang2019particlelike}. The probe beam \bm{$E_1$} from a second laser, focused by a lens onto the front surface of the atomic vapor cell, propagates through the cell and is collected by another lens with the same focal length. Then it is  detected by a CCD camera placed at the collection lens' back focal plane, which shows the Fourier image presenting the momentum space of the probe beam. The phase measurements are made by homodyne detection, i.e. by interfering the probe beam with a reference beam from the same laser at the position of the CCD, where the fork-like structure of the interference stripes indicates the formation of vortices. The susceptibility experienced by \bm{$E_1$} is determined by the frequency detunings of \bm{$E_1$} (denoted as $\Delta_1$) and \bm{$E_2$} (denoted as $\Delta_2$) via the EIT effect in a three-level atomic configuration [Fig.~\ref{fig_exp_setup}(b)]. When the two-photon detuning $\Delta_1$-$\Delta_2$ is set to result in higher (resp. lower) refractive index at the bright sites of the hexagonal pattern, the bright (resp. dark) sites forms a hexagonal (resp. honeycomb) photonic lattice. Figures~\ref{fig_exp_setup}(c) and (d) show the simulated spatial distribution of refractive index of the honeycomb and hexagonal lattices, respectively. The Gaussian probe beam is focused to a spot size comparable to that of a honeycomb lattice site. During propagating along $z$ in the 2D lattice, which can be considered as a matrix of coupled waveguides, the probe beam expands within the $x$-$y$ plane and interacts with the 2D lattice to form vortices. 

\begin{figure}
\includegraphics[width=0.45\textwidth]{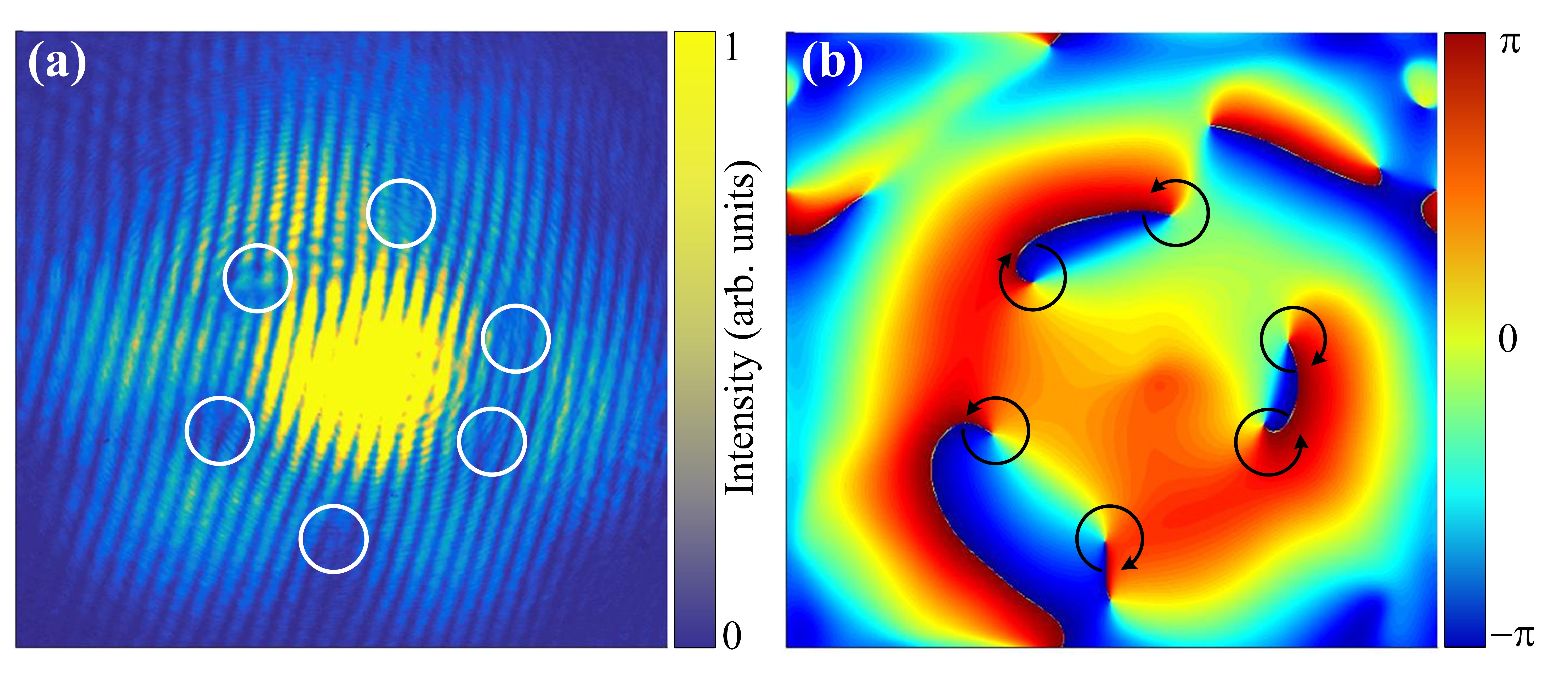}\\
\caption{\label{fig_exp} Momentum-space vortex generation in the honeycomb lattice with the two-photon detuning being 20~MHz. (a) Experimentally measured momentum-space image interference with the reference beam. The dislocations in fringes correspond to the vortices (marked by white circles). (b) The corresponding phase pattern extracted from the interference image. Black arrows show the rotation direction.}
\end{figure}

We fist set the frequency detunings as $\Delta_1=-80$~MHz and $\Delta_2=-100$~MHz which yields a honeycomb lattice in the Rb cell with a lattice constant of $\sim$ 76 $\mu$m, as shown in Fig.~\ref{fig_exp_setup}(c). The probe beam is centered at one of the honeycomb lattice sites. After propagating through the Rb cell, the interference pattern between the momentum space image and the reference beam shows six evenly distributed fork-like features, indicating the presence of six vortices, as shown in Figs.~\ref{fig_exp}(a) and (b). The vortices are associated with Dirac points and their signs depend on the neighboring valley type, $K$ or $K'$. One could naturally interpret these vortices and anti-vortices as a result of the pseudospin conversion with a geometric phase \cite{liu2020universal}. The probe beam, which is focused to a small spot, contains a large range of incident angles (determined by the lens numerical aperture) that corresponds to a continuum of photon momenta. These high-angle probe components can generate vortices detected at the corresponding positions in the momentum space, providing an easy technology of generating multiple and spatially well-separated vortices using non-structured light. Additionally, we find that creation of multiple vortices is very sensitive to the initial position (the site with higher or lower refractive index) of the focused probe beam on the honeycomb lattice, indicating that the process occurs only under special situations highly restricted by symmetry. To fully understand this, we perform a quantitative theoretical analysis of the beam propagation in optical lattices and interpretation of symmetry effects.

The field profile in the cell can be written in general form as
\begin{equation}
    \vec{E}(x,y,z,t)=\vec{E_0}a(x,y,z)e^{i(k_0nz-\omega t)},
\end{equation} 
where $\omega$ is the frequency of the laser beam, $n$ is the refraction coefficient, $k_0$ is the wave vector of light in the vacuum, $\vec{E_0}$ is the amplitude vector, and $a$ defines the slowly varying profile of electric field.

In the paraxial approximation $\partial^2 a/\partial z^2\ll k_0\partial a/\partial z$ the electric field envelope $a$ obeys the parabolic partial differential equation
\begin{equation}
\label{eq_parax}
  i\partial_z a = \left[ - \frac{\Delta}{{2{k_0}{n_0}}} - k_0^2\left( {{n^2} - n_0^2} \right) \right] a,
\end{equation}
where $\Delta = \partial_x^2+\partial_y^2$, $n=n(x,y)$ is the laterally varying part of the refractive index due to the optical modulation with the coupling field, $n_0$=1 is the background refraction index.

Equation \eqref{eq_parax} is equivalent to the time-dependent Schr\"odinger equation with $z$ coordinate playing the role of time, with the mass given by $m=\hbar k_0 n_0/c$, and the potential energy $U(x,y)$ determined by the spatial variation of the refraction index: $U(x,y)=-\hbar c k_0^2(n^2-n_0^2)$, where $c$ is the speed of light. The initial condition for the envelope field $a$ is governed by the Gaussian profile of the probe beam entering the vapor cell (at $z=0$). The time-dependent Eq.~\eqref{eq_parax}  with the honeycomb lattice potential was solved numerically in the general case of spatially varying effective potential lattice $U(x,y)$ and in the tight-binding approximation.

\begin{figure}
\includegraphics[width=0.48\textwidth]{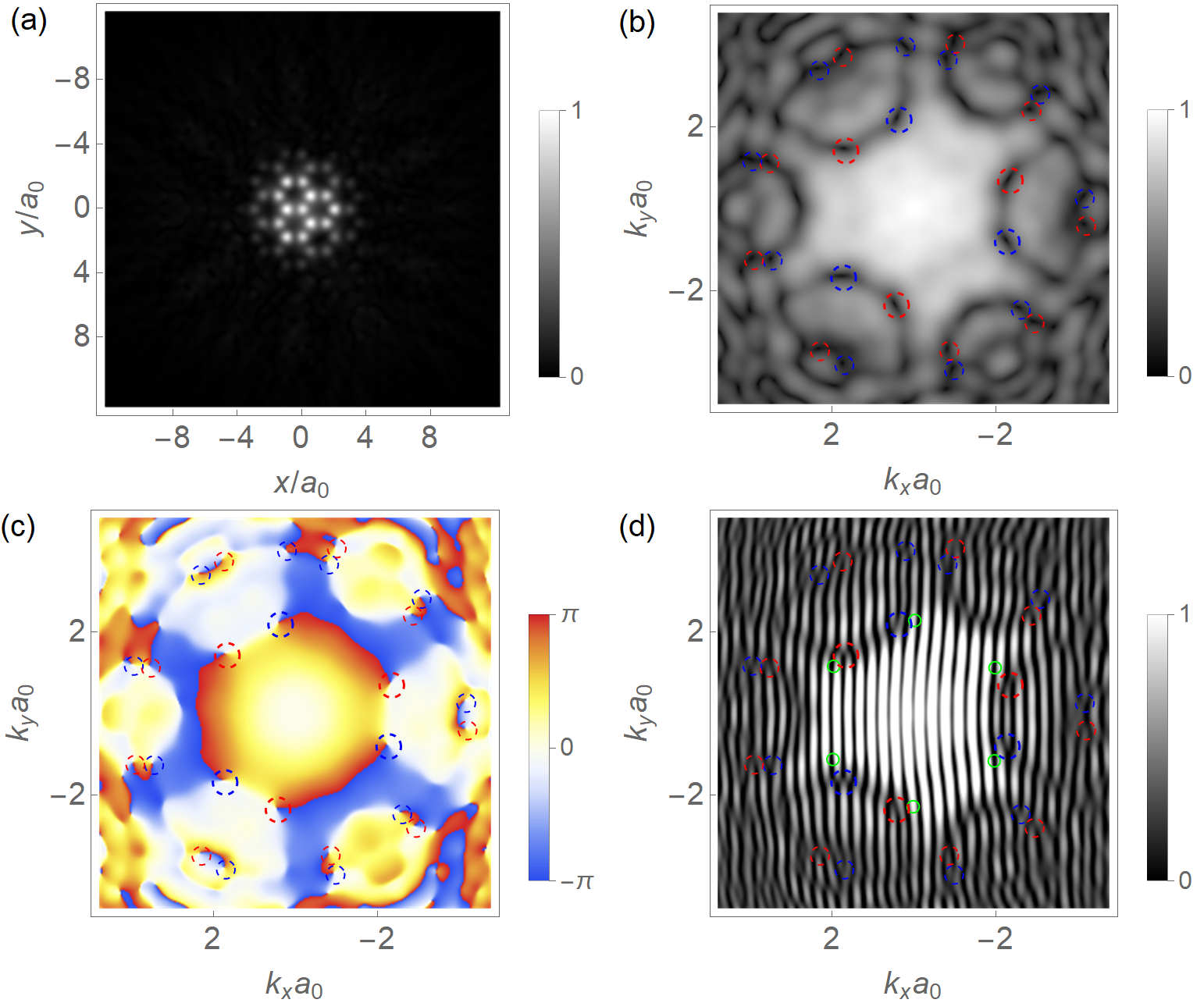}
\includegraphics[width=0.48\textwidth]{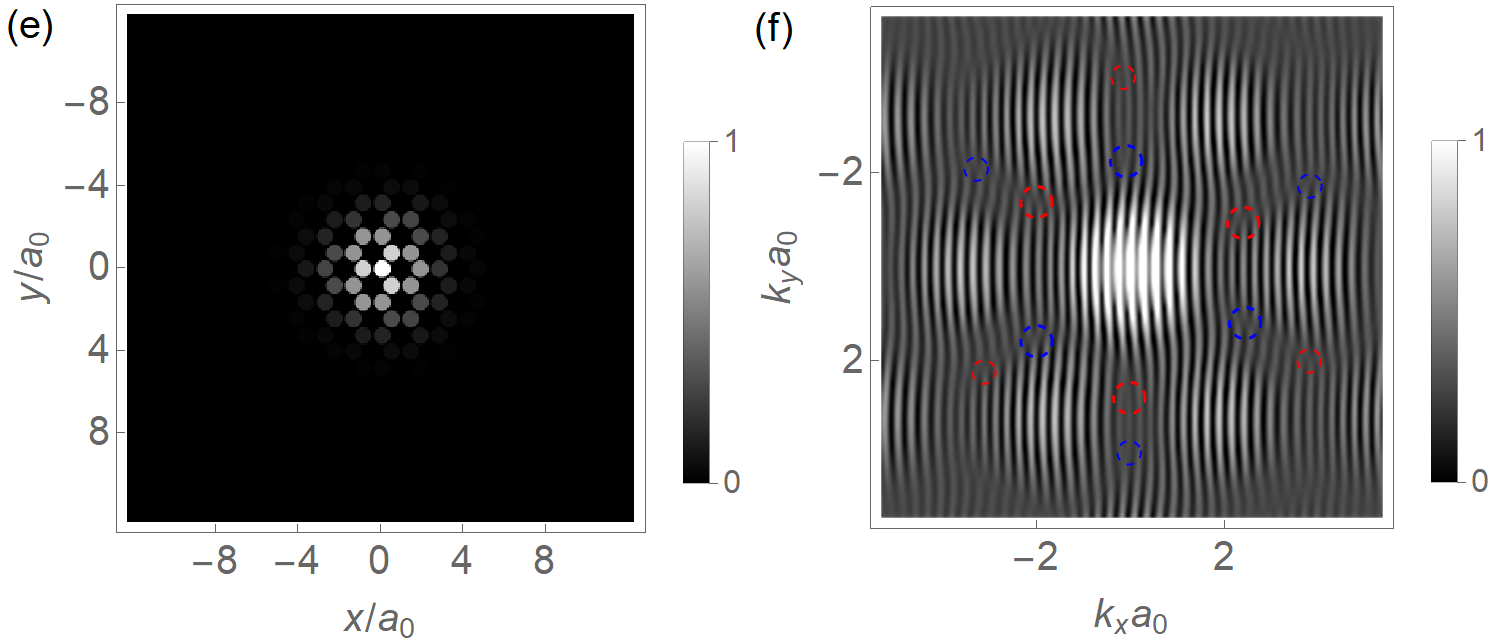}\\
\caption{\label{fig_schr_gra_tri_main} Numerical solution of the Schr\"odinger equation showing wave packet expansion in photonic graphene starting from the excitation of a single site.
(a) The probability distribution in real space.
(b) The probability distribution in momentum space.
(c) The reciprocal-space phase image of the WF corresponding to panel (b). The left and right vortices are marked with blue and red circles, respectively.
(d) The reciprocal-space WF interference with a plane wave. Green circles indicate the Dirac points.
The snapshot shows that the WF has the $C_{3\rm v}$ symmetry. Panels (e) and (f) show the results of tight-binding calculation and correspond to panels (a) and (d), respectively.
 }
\end{figure} 

Emergence of the momentum-space quantized vortices was first demonstrated with the full numerical solution of Eq.~\eqref{eq_parax}. The initial WF, which was defined as a narrow Gaussian spatial profile with the width comparable to the site size, was centered on one of honeycomb lattice sites as in experiment. In this excitation configuration, the real-space WF of the probe beam exhibits $C_{3\rm v}$ symmetry during time evolution, as shown in Fig.~\ref{fig_schr_gra_tri_main}(a).  The calculated momentum-space profiles, as depicted in Figs.~\ref{fig_schr_gra_tri_main}(b)-(d) for probability density, phase and interference patterns, show the quantum vortices in the vicinity of Dirac points robustly determined by the automatic vortex detection algorithm~\cite{koniakhin20202d}, agreeing very well with the experimental measurements. The real part of the WF Fourier image has sixfold rotation symmetry, whereas the imaginary part obeys the $C_{3\rm v}$ symmetry in accordance with Refs.~\cite{boguslawski2011increasing,koniakhin2014ratchet,dutreix2019measuring}. The obtained result is similar to Fig.~1a from Ref. \cite{boguslawski2011increasing}, with the exchanged real and momentum space.

The time-dependent tight-binding-type model (TBM) calculations on the honeycomb lattice ascertain the robustness of the results and their insensitivity to the particular shape of the site potential. Figures~\ref{fig_schr_gra_tri_main}(e) and (f) show the result of the time evolution of the WF for the initial condition with the single site excited and zero WF at others, which corresponds to the numerical solution of full Schr\"odinger equation and the experimental design. The results of the TBM were then converted into real-space field using the assumption of uniform WF distribution at each site for direct comparison of the two models. Spline interpolation of the WF using the TBM results was also employed, both approaches showed qualitatively similar results, confirming the validity of the model. Figure~\ref{fig_schr_gra_tri_main}(e) shows the probability density in real space. Figure~\ref{fig_schr_gra_tri_main}(f) depicts the interference pattern for the WF in momentum space. The Brillouin zone replicas are visible better than in Schr\"odinger equation.

\begin{figure}[h]
\includegraphics[width=0.48\textwidth]{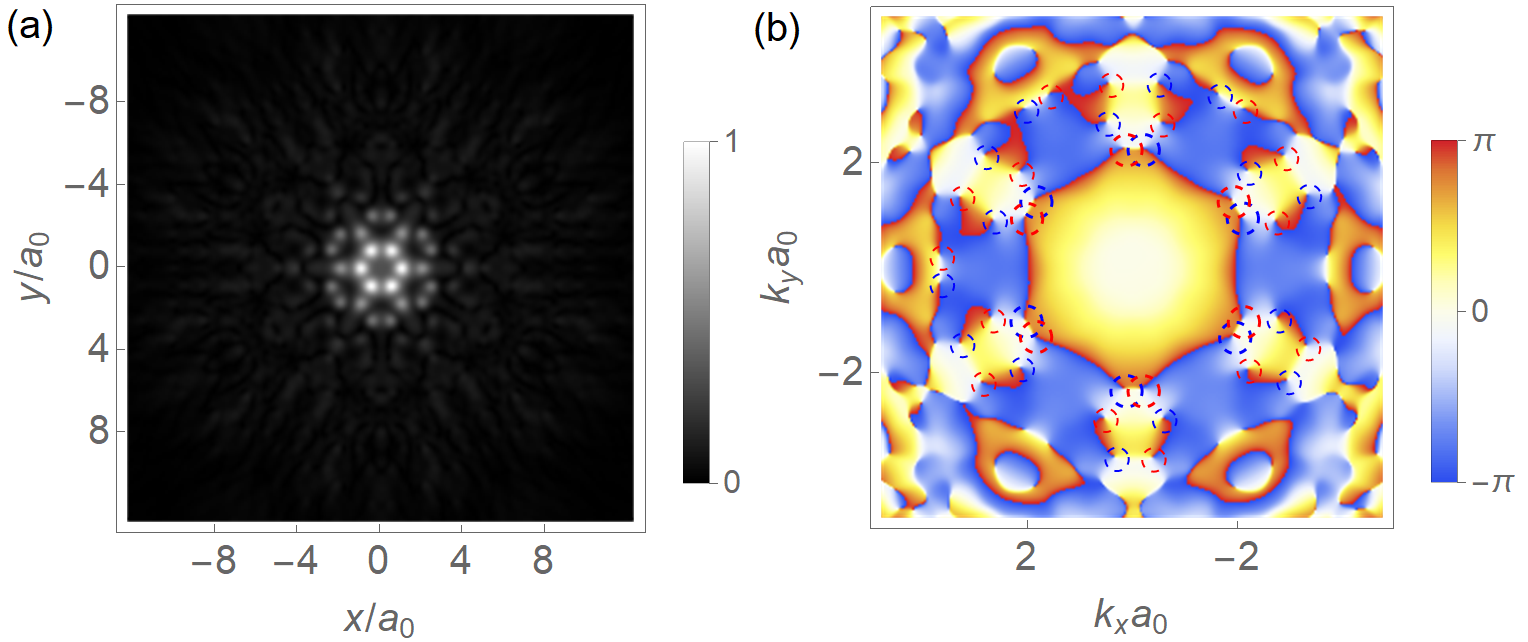}\\
\includegraphics[width=0.48\textwidth]{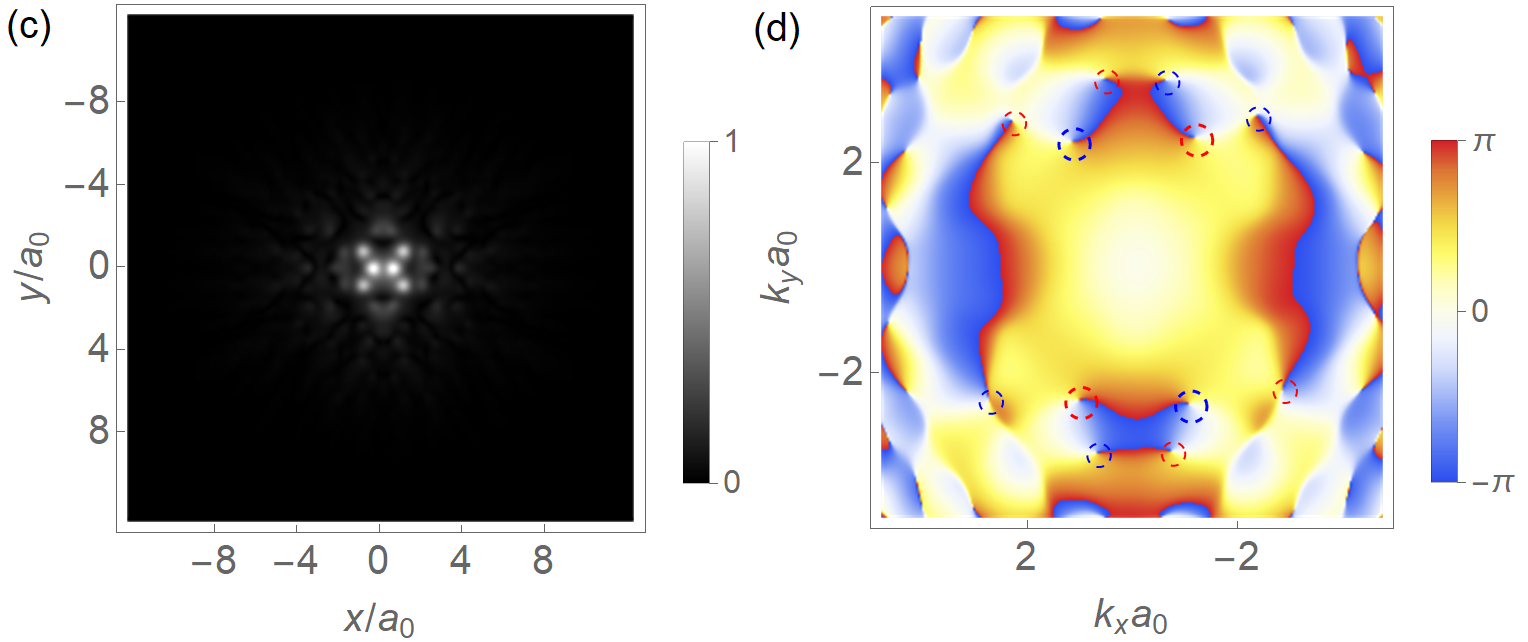}\\
\includegraphics[width=0.48\textwidth]{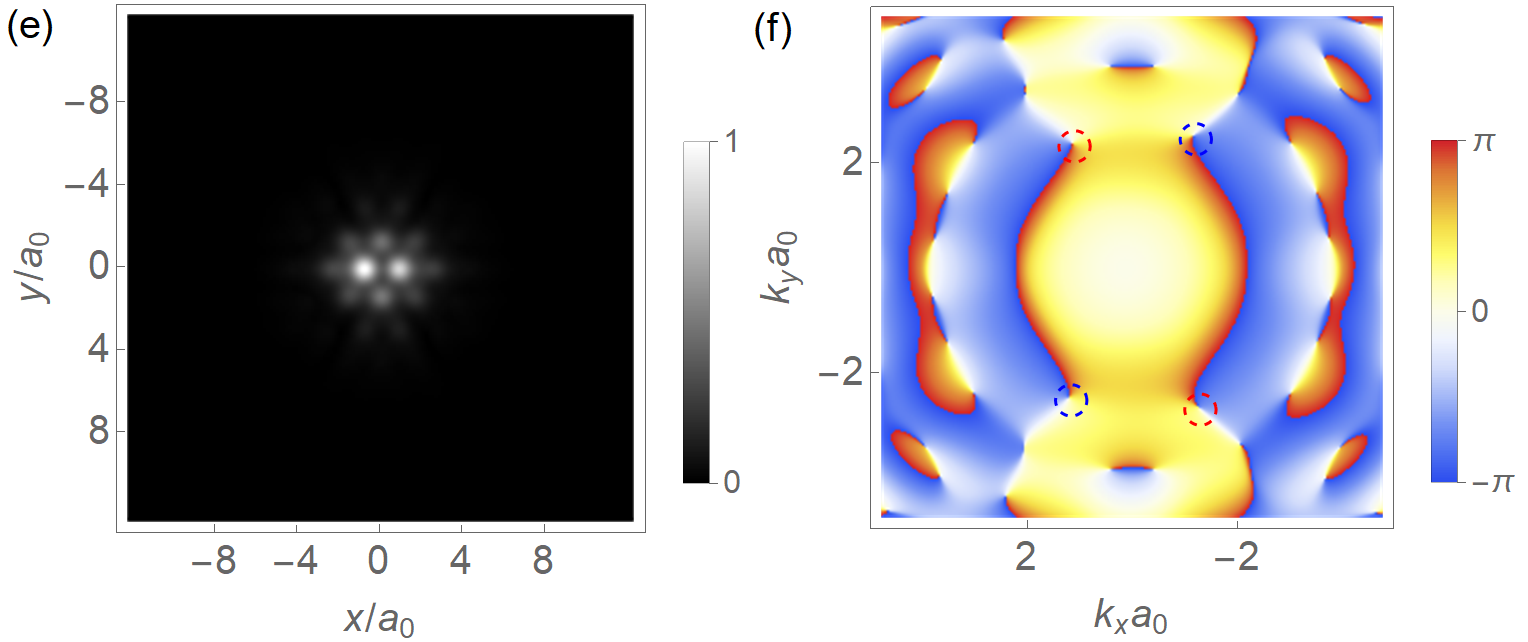}
\caption{\label{fig_hex_hex}  Numerical solution of the Schr\"odinger equation showing wave packet expansion in photonic honeycomb and hexagonal lattices with various symmetries of the WFs. (a,b) Honeycomb lattice, excitation of a single ``benzene ring'' ($C_{6\rm v}$): real-space probability (a)  and momentum space phase (b). (c,d) The same for a honeycomb lattice, excitation between two sites ($C_{2\rm v}$ symmetry).(e,f) The same for a topologically trivial hexagonal lattice with $C_{2\rm v}$-excitation.}
\end{figure}

Another types of the WF symmetry that can be realized in the honeycomb lattice are the $C_{6\rm v}$ and $C_{2\rm v}$ symmetries. Figures~\ref{fig_hex_hex}(a) and (b) show the result of the time evolution of WF and its Fourier transform for hexagonally-symmetric initial excitation. For a  $C_{6\rm v}$-object, both real and imaginary parts of the momentum-space WF inherit its symmetry, which forbids the presence of single opposite-signed vortices near the Dirac points due to the presence of mirror reflection symmetry. Figures~\ref{fig_hex_hex}(c) and (d) demonstrate the vortices in the momentum space for initial excitation between the two sites giving the dimer-like structure. In this case, the vortices in momentum space are positioned at the rectangle corners.

Along with the symmetry, radial distribution of the signal also plays an important role for vortex generation. Consider the  WF of $C_{3\rm v}$ symmetry on graphene lattice, where the six-vortex array can be expected in momentum space. The most simple WF obeying these conditions is a tetramer, formed of one central site and three neighbors with the WFs $\Psi=1$ and $\Psi=ae^{i\varphi}$, respectively. Generation of vortices depends on the parameters $(a,\varphi)$, but they always exist for $a>1$, see Fig.~\ref{fig_PD}.

As the honeycomb lattice originates the momentum-space valleys that can be related to topologically nontrivial nature, one important question was whether the generation of multiple vortices from lattices has to be linked to topology and momentum-space valleys. The similar WF of $C_{2\rm v}$ symmetry can be potentially obtained in the trivial hexagonal lattice. Figures~\ref{fig_hex_hex}(e) and (f) show time evolution of the WF in real space and phase image in momentum space, which exhibits rectangular vortex patterns very similar to Fig.~\ref{fig_hex_hex}(d).

\begin{figure}[h]
\includegraphics[width=0.45\textwidth]{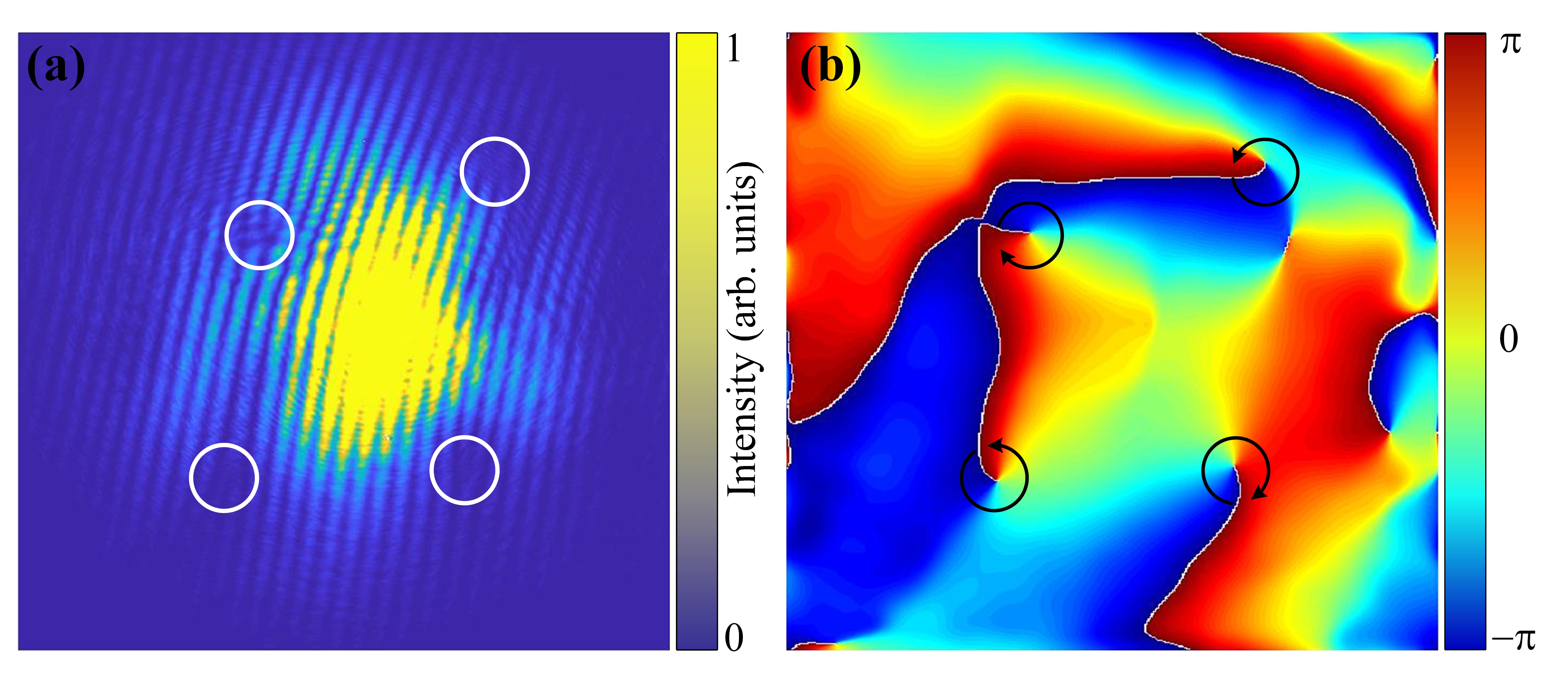}\\
\caption{\label{fig_exp_hex_rect} Momentum-space vortex generation in the hexagonal lattice with the two-photon detuning being -90~MHz. (a) Observed momentum-space image interfered with reference beam. The fork-like fringes correspond to the vortices (white circles). (b) The corresponding extracted phase pattern (black arrows -- the rotation direction).}
\end{figure}

To test the corresponding configuration experimentally, we have established the required topologically trivial hexagonal lattice with a lattice constant of $\sim$131$\mu$m, by setting the frequency detunings to $\Delta_1=-190$~MHz and $\Delta_2=-100$~MHz, as simulated in Fig.~\ref{fig_exp_setup}(d). The probe beam was focused at the middle point between two lattice sites. Four vortices, composed of two vortex and anti-vortex pairs, appear at the corners of a slightly skewed rectangle, as shown in Fig.~\ref{fig_exp_hex_rect}. This indicates that the generation of multiple vortices can still occur in topologically trivial structures, advocating that the symmetry plays the main role instead of topology.

Thus one can conclude that topological nature of lattice (and even an infinite lattice itself) does not play a key role for generation of vortices in momentum space, but the symmetry and the radial profile of the WF do. Fourier transform maps the symmetry of the WF in real space to the momentum space. At the same time, the photonic lattices are a very flexible and simple to implement source of symmetries that will be inherited by the evolving wave packet. In addition, the lattice defines the size of the wave packet in real space, and also consequently defines the profile of the wave packet in momentum space, including positions of vortices if they exist. The particular symmetry of the WF as well as its radial distribution can be shaped by position and profile of the excitation beam. We benefit from advanced application of discrete non-diffracting beams \cite{boguslawski2011increasing} by usage of real and momentum space equivalence and control of the signal symmetry. Even the simplest Gaussian beams carrying no OAM allow flexible engineering of the WF profile and realization of momentum-space vortices with multiple configurations. 

The symmetries were always understood as being absolutely fundamental in physics \cite{Landau1}. While in previous decades  the topological invariants have appeared in solid state physics as something seeming to go beyond  symmetries, recent works start to reestablish their relation to symmetries via the so-called symmetry indicators \cite{fu2007topological,po2017symmetry,song2018quantitative,tang2019efficient}, allowing to determine the topology from the eigenstates at high symmetry points. Our work makes an important step in this direction, demonstrating that the formation of topological defects such as vortices in periodic potentials is determined by the lattice symmetries, and not by its topology.

In conclusion, we have shown the presence of vortices in momentum space aligned in the vicinity of Dirac points in photonic graphene based on the atomic vapor cell with spatially modulated susceptibility after the scattering of narrow probe beam. This configuration is an analogue of the wave packet evolution in honeycomb lattice in the framework of Schr\"odinger equation with single-site initial excitation. In the honeycomb lattice, $C_{3\rm v}$ symmetry of the wave packet supports, wherease $C_{6\rm v}$ symmetry cancels the presence of vortices near the Dirac points. In the non-topological hexagonal lattice, vortices form a rectangle pattern with the lower (mirror reflection) symmetry of the spreading wave packet. These results show that the creation of multiple vortices near the Dirac points is strongly linked with the wave packet symmetry, but not necessarily with the topological nature of the lattice.  The resulting spatial pattern of vortices is determined by the product of the probe light and the lattice symmetries. Our results provide a novel method to generate well-distinguished vortex arrays in the simplest manner that requires no structured probe light, and reveal the basic rules of how to achieve the desired spatial distribution of vortices by designing the initial symmetry of the lattice and of the probe beam.

\begin{acknowledgments}
This work was supported by National Key R\&D Program of China (2018YFA0307500), the Key Scientific and Technological Innovation Team of Shaanxi Province (2021TD-56), National Natural Science Foundation of China (12074303, 62022066, 12074306, 11804267). S.V.K. acknowledges the IBS Young Scientist Fellowship (IBS-R024-Y3-2021). E.C. acknowledges Basis foundation (Grant 21-1-3-30-1). D.D.S. and G.M. acknowledge the support of the European Union's Horizon 2020 program, through a FET Open research and innovation action under the grant agreement No. 964770 (TopoLight), project ANR Labex GaNEXT (ANR-11-LABX-0014), and of the ANR program "Investissements d'Avenir" through the IDEX-ISITE initiative 16-IDEX-0001 (CAP 20-25). 
\end{acknowledgments}

%\appendix
%\counterwithin{figure}{section}
%\section{Supplemental materials. Phase diagram for minimal $C_{3\rm v}$ symmetry wave function}
\renewcommand{\thefigure}{S1}
\begin{figure}[h]
\begin{center}
\includegraphics[width=0.48\textwidth]{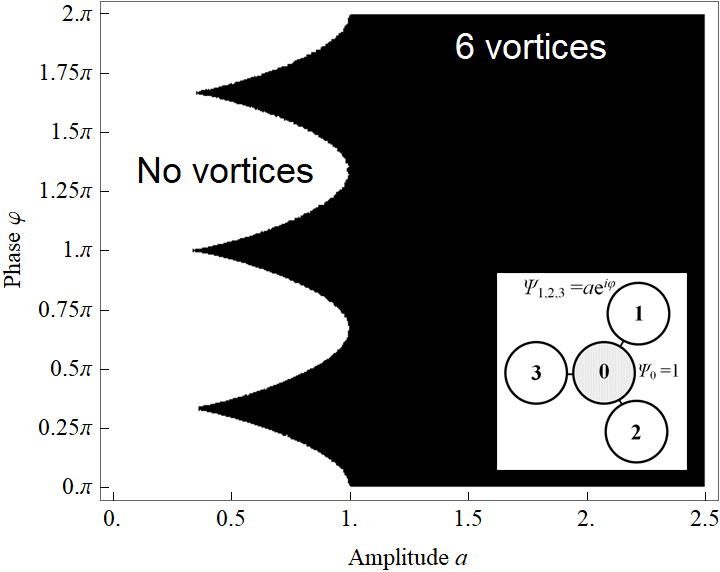} 
\end{center}
\caption{\label{fig_PD} \underline{Supplemental material}. Formation of the regular 6 vortices in momentum space (as in Fig. 3) for the tetramer of $C_{3\rm v}$ symmetry. The central site has the wave function $\Psi_0=1$ and three neighbors have the wave function $\Psi_{1,2,3}=ae^{i\varphi}$. Black color at the phase diagram of parameters $a$ and $\phi$ corresponds to the formation of vortices and white color is for their absence.}
\end{figure}

\bibliography{bib}

\end{document}